\documentclass[prc,aps,eqsecnum,twocolumn,floatfix,nofootinbib]{revtex4-1}
\usepackage{ulem}  
\usepackage{dcolumn}
\usepackage{bm}
\usepackage{color}
\usepackage{amssymb}
\usepackage{amsmath}
\usepackage{graphicx}
\usepackage{amsfonts}
\usepackage{slashed}
\usepackage{pstricks}
\usepackage{float}
\usepackage{hyperref}
\usepackage{array}
\usepackage{epsf}
\usepackage{soul}
\usepackage{siunitx}
\usepackage{multirow}
\newcommand{\bra}{\langle}
\newcommand{\ket}{\rangle}
\newcommand{\Ht}{{}^3{\rm H}}
\newcommand{\Het}{{}^3{\rm He}}
\newcommand{\m}{\phantom{-}}
\newcommand{\z}{\phantom{0}}
\allowdisplaybreaks
\begin{document}
\title{Magnetic structure of few-nucleon systems at high momentum transfers
in a $\chi$EFT approach}
\author{A.\ Gnech$^{\rm a}$ and R.\ Schiavilla$^{\rm a,b}$, }
\affiliation{
$^{\rm a}$\mbox{Theory Center, Jefferson Lab, Newport News, Virginia 23606, USA}\\
$^{\rm b}$\mbox{Department of Physics, Old Dominion University, Norfolk, Virginia 23529, USA}\\
}
\date{\today}
\begin{abstract}
The five low-energy constants (LECs) in the electromagnetic current
derived in chiral effective field theory ($\chi$EFT) up to one loop are
determined by a simultaneous fit to the $A\,$=$\,2$--3 nuclei magnetic moments
and to the deuteron magnetic form factor and threshold electrodisintegration
at backward angles over a wide range of momentum transfers.  The resulting
parametrization then yields predictions for the
$^3$He/$^3$H magnetic form factors in excellent accord with the experimental
values for momentum transfers ranging up to  $\approx 0.8$ GeV/c, 
beyond the expected regime of validity of the $\chi$EFT approach.
The calculations are based on last-generation two-nucleon interactions
including high orders in the chiral expansion and derived by
Entem, Macheleidt, and Nosyk [Phys.\ Rev.\ C {\bf 96}, 024004 (2017)] and
by Piarulli {\it et al.} [Phys.\ Rev.\ C {\bf 94}, 054007 (2016)], using different
$\chi$EFT formulations.  In the $A\,$=$\,3$ calculations, (chiral) three-nucleon
interactions are also accounted for.  The model dependence resulting from these
different formulations of the interactions is found to be mild for momentum transfer
below $\approx0.8$ GeV/c. An analysis of the convergence of the chiral expansion
is also provided.
 \end{abstract}
 
\index{}\maketitle
\section{Introduction and conclusions}
\label{sec:intro}
The nuclear electromagnetic current derived in chiral effective field theory ($\chi$EFT)
at one loop is characterized by five low-energy constants (LECs).  Three of these
LECs enter a subleading one-pion-exchange (OPE) term, while the remaining two are
associated with contact terms induced by non-minimal coupling to the electromagnetic
field.  Resonance saturation arguments can be used~\cite{Pastore:2009}
(and have been used~\cite{Piarulli:2013,Schiavilla:2019}) to relate the two LECs in
the isovector component of the subleading OPE current to the $N$-to-$\Delta$ transition axial
coupling and transition magnetic moment.
The remaining LECs have been determined by fitting experimentally known few-nucleon
observables at low energy and momentum transfers, such as the magnetic moments of deuteron, $^3$H, and
$^3$He~\cite{Piarulli:2013,Schiavilla:2019} and/or the deuteron magnetic
form factor~\cite{Koelling:2012} and/or the cross section for radiative
capture of thermal neutrons on protons~\cite{Piarulli:2013}.
 
In the present work we adopt a different strategy for constraining these LECs:
(i) we do not invoke resonance saturation, and (ii) we determine the complete
set of five LECs by a simultaneous fit to the magnetic moments of the $A\,$=$\,2$--3
nuclei, and the deuteron magnetic form factor and threshold electrodisintegration
cross section at backward angles over a wide range of momentum transfers.
It turns out that these five LECs provide enough flexibility to allow us to reproduce
well all these observables in a region of momentum transfers that extends above 4 fm$^{-1}$.
That a satisfactory fit of these high momentum transfer data is possible, was not
anticipated.  It suggests that the present parametrization of the current is robust
in kinematical regimes (momentum transfers of the
order of 0.8 GeV/c) outside the limits of validity of the $\chi$EFT expansion.
This conclusion is further corroborated by the excellent agreement between the
measured and predicted magnetic form factors of $^3$H/$^3$He for
momentum transfers up to 0.8 GeV/c, including partially their diffraction regions.
Previous parametrizations of the $\chi$EFT
current, using resonance saturation arguments and based on fits of static
magnetic properties of deuteron and trinucleons, had failed to provide a good description of these regions~\cite{Piarulli:2013,Schiavilla:2019}.

The present paper is organized as follows.  In the next section
we list the expressions for the $\chi$EFT current up to one loop.
These are well known by now, but are reported here for completeness
and to facilitate the ensuing discussions.  In Sec.~\ref{sec:fit}
we detail the determination of the LECs and associated uncertainties,
and in Sec.~\ref{sec:res} we present a comparison between the
measured trinucleon magnetic form factors and those predicted by the present
parametrization of the $\chi$EFT current. We also show in that
section that the (crucial) two-body
terms in the current yield contributions in the deuteron threshold electrodisintegration
and  $A\,$=$\,3$ magnetic form factors that are proportional to each other.
The implications of this fact are discussed.

\section{EM currents up to one loop}
\label{sec:c-cnt}
Independent derivations of nuclear electromagnetic current and charge operators up to one loop in $\chi$EFT
have been carried out within a variety of different formalisms and in a number of
papers~\cite{Park:1996,Pastore:2008,Pastore:2009,Koelling:2009,Pastore:2011,Koelling:2011,Piarulli:2013}
in the past three decades or so. For clarity of presentation and for later reference, we report below
the explicit expressions for the electromagnetic current operators of interest in the present work.  We adopt the notation of Refs.~\cite{Pastore:2009,Piarulli:2013};
in particular, we denote the generic low-momentum scale with $Q$ and the
momentum due to the external electromagnetic field with ${\bf q}$, and define
\begin{eqnarray}
{\bf k}_i&=&{\bf p}_i^\prime-{\bf p}_i \ ,\qquad {\bf K}_i=\left({\bf p}_i^\prime+{\bf p}_i\right)/2 \ ,\\
{\bf k}&=&\left({\bf k}_1-{\bf k}_2\right)/2 \ , \qquad {\bf K}= {\bf K}_1+{\bf K}_2 \ ,
\label{eq:ppp}
\end{eqnarray}
where ${\bf p}_i$ (${\bf p}_i^\prime$) is the initial (final) momentum of nucleon $i$.  We
further note that in the ${\bf j}^{(n)}$ expressions reported below
the superscript $n$ specifies the order $e\, Q^n$ in the power counting
(in a two-nucleon system).
The lowest-order (LO) contribution ${\bf j}^{(-2)}$ consists of the single-nucleon current
\begin{eqnarray}
\label{eq:jlo}
{\bf j}^{(-2)}_{{\rm NR}}({\bf q})&=&\frac{e}{2\, m_N}
\left[ \,2\, e_{N,1}(q_\mu^2) \, {\bf K}_1 
+i\,\mu_{N,1}(q_\mu^2)\, {\bm \sigma}_1\times {\bf q }\,\right] \nonumber\\
&&\times\, \delta({\bf p}^\prime_2-{\bf p}_2)+
 1 \rightleftharpoons 2\ ,
 \end{eqnarray}
where $m_N$ is the nucleon mass,
${\bf q}\,$=$\,{\bf k}_i$ with $i\,$=$\,1$ or 2 (the $\delta$-function enforcing
overall momentum conservation ${\bf q}\,$=$\,{\bf k}_1$ has been dropped for brevity here and
in the following),
\begin{eqnarray}
e_{N,i}(q_\mu^2) &=& \left[G_E^S(q_\mu^2)+G_E^V(q_\mu^2)\, \tau_{i,z}\right]/2\ , \nonumber \\ 
 \mu_{N,i}(q_\mu^2) &=&\left[ G_M^S(q_\mu^2)+G_M^V(q_\mu^2)\, \tau_{i,z}\right]/2 \ ,
\label{eq:ekm}
\end{eqnarray}
and $G^{S/V}_E$ and $G^{S/V}_M$ denote the isoscalar/isovector
combinations of the proton ($p$) and neutron ($n$) electric ($E$) and magnetic ($M$)
form factors
\begin{equation}
G_{E/M}^{S/V}(q_\mu^2)=G_{E/M}^p(q_\mu^2)+\!/\!-G_{E/M}^n(q_\mu^2) \ .
\end{equation}

The power counting $e\, Q^{-2}$ of this current
results from the product of a factor $e\, Q$ due to
the coupling of the external electromagnetic field to the individual nucleons, and
the factor $Q^{-3}$ from the momentum $\delta$-function entering this type of
disconnected contributions.  Of course, such a counting ignores the fact that the
nucleon form factors themselves also have a power series expansion in $Q$.
Here, they are taken from fits to elastic electron scattering data off the proton
and deuteron rather than derived consistently in chiral perturbation theory ($\chi$PT)~\cite{Kubis01};
specifically, we utilize the dipole parametrization with
\begin{eqnarray}
\nonumber
G_E^p(q_\mu^2)\! &=&\! G_D(q_\mu^2)   \ , \,\,\,\,\,
G_E^n(q_\mu^2)\! =\!-\mu^n\, \frac{q_\mu^2}{4\, m_N^2} \frac{G_D(q_\mu^2)}{1+q_\mu^2/m_N^2} \ ,\\
 G_M^p(q_\mu^2) &=& \mu^p G_D(q_\mu^2) \ ,\,\,\,\,\, G_M^n(q_\mu^2) = \mu^n G_D(q_\mu^2)\ ,
\nonumber
\end{eqnarray}
where $\mu^p$ and $\mu^n$ are, respectively, the proton and neutron magnetic moments, and
\begin{equation}
G_D(q_\mu^2)=\left(1+q_\mu^2/\Lambda^2\right)^{-2} \ ,
\end{equation}
with $\Lambda\,$=$\,0.83$ GeV. We take these form factors as functions of the four-momentum
transfer $q_\mu^2\,$=$\,q^2-\omega^2$, where $\omega$ is the energy transfer.  We also note
that the calculation presented below are carried out in the laboratory frame.

At order $n=\! -1$ (NLO) there is a one-pion exchange (OPE) contribution that reads
\begin{eqnarray}
 {\bf j}^{(-1)}_\pi({\bf q})\!&=&\! -i\, e\frac{g^2_A}{4f^2_\pi}\,G^V_E(q_\mu^2)\,
 ({\bm \tau}_1 \times {\bm \tau}_2)_z 
 \bigg( {\bm \sigma}_1 -{\bf k}_1\,\frac{{\bm \sigma}_1\cdot {\bf k}_1} {\omega_{k_1}^2}  \bigg)\nonumber\\
&&\times \frac{{\bm \sigma}_2\cdot {\bf k}_2}{\omega^2_{k_2}} + 1 \rightleftharpoons 2 \ ,
 \label{eq:nlo1} 
 \end{eqnarray}
where $g_A$ is the nucleon axial coupling constant ($g_A\,$=$\, 1.29$),
$f_\pi$ is the pion decay amplitude ($f_\pi\,$=$\, 92.4$ MeV) and 
we have defined $\omega_k^2=k^2+m_\pi^2$, with $m_\pi$ being the 
pion mass.  The inclusion of the isovector electric form factor $G_E^V(q_\mu^2)$ in ${\bf j}^{(-1)}$
can be justified on the basis of the continuity equation, see Ref.~\cite{Piarulli:2013}.

Relativistic corrections to the leading order one-body current
operators enter at $n=0$ (denoted as N2LO), and are given by
\begin{eqnarray}
\label{eq:j1rc}
 \!\!\!\!{\bf j}_{\rm RC}^{(0)}({\bf q})&=&-\frac{e}{8 \, m_N^3}
 e_{N,1}(q_\mu^2) \, \Big[ 
2\, \left( K_1^2 +q^2/4 \right)  \big( 2\, {\bf K}_1 \nonumber\\
\!\!\!\!&&\!\!\!\!+i\, {\bm \sigma}_1\times {\bf q } \big) 
+ {\bf K}_1\cdot {\bf q}\, \left({\bf q} +2\, i\, {\bm \sigma}_1\times {\bf K }_1 \right)\Big]\nonumber\\ 
\!\!\!\!&&\!\!\!\!- \frac{i\,e}{8 \, m_N^3}
\left[\, \mu_{N,1}(q_\mu^2)-e_{N,1}(q_\mu^2)\, \right]
 \Big[ {\bf K}_1\cdot {\bf q}\\
\!\!\!\!&&\!\!\!\!\times
 \big( 4\, {\bm \sigma}_1\times {\bf K}_1-i\, {\bf q}\big) 
 - \left(  2\, i\, {\bf K}_1 -{\bm \sigma}_1\times {\bf q} \right)\, q^2/2 \nonumber\\
\!\!\!\! &&\!\!\!\!  +2\, \left({\bf K}_1\times {\bf q}\right)
 \, {\bm \sigma}_1\cdot {\bf K}_1 \Big] \, \delta({\bf p}^\prime_2-{\bf p}_2) + 1 \rightleftharpoons 2  \ .
 \nonumber
\label{eq:r1rc}
\end{eqnarray}
In the calculations of electromagnetic observables to follow, we also utilize chiral $2N$ and $3N$ interactions
which retain explicitly $\Delta$-isobar degrees of freedom, the NV models of
Refs.~\cite{Piarulli:2015,Piarulli:2016,Piarulli:2018}.   In these instances, we account
for the N2LO currents originating from explicit $\Delta$ intermediate states, given by~\cite{Pastore:2008}
\begin{eqnarray}
\label{eq:cdlta}
{\bf j}^{(0)}_{\Delta}({\bf q})&=&i\, e\, \frac{g_A\,h_A}{18\,m_{\Delta N}f_\pi^2} G_{\gamma N\Delta}(q_\mu^2)\,
\frac{{\bm \sigma}_2 \cdot {\bf k}_2}{\omega_{k_2}^2}\,
\bigg[  4\, \tau_{2,z} \,{\bf k}_2 \nonumber \\
&&- ({\bm \tau}_1\times{\bm \tau}_2)_z\, {\bm \sigma}_1\times {\bf k}_2  \bigg] \times {\bf q} + 1\rightleftharpoons 2 \ ,
\end{eqnarray}
where $m_{\Delta N}$ is the $\Delta$-nucleon mass difference ($m_{\Delta N}\,$=$\, 293$ MeV),
$h_A$ and $G_{\gamma N\Delta}$ are, respectively, the $N$-to-$\Delta$ transition axial coupling
constant ($h_A\,$=$\,2.74$) and transition electromagnetic form factor.  The latter is
parametrized as
\begin{equation}
G_{\gamma N \Delta}(q_\mu^2)= \frac{\mu_{\gamma N \Delta} }
{( 1+q_\mu^2/\Lambda_{\Delta,1}^2 )^2
\sqrt{1+q_\mu^2/\Lambda_{\Delta,2}^2} } \ ,
\end{equation}
where $\mu_{\gamma N \Delta}$---the transition magnetic moment---is taken to be $3\,\mu_N$ from
an analysis of $\gamma N$ data
in the $\Delta$-resonance region~\cite{Carlson86}.  This
analysis also gives $\Lambda_{\Delta,1}$=0.84 GeV and
$\Lambda_{\Delta,2}$=1.2 GeV.  

The currents at order $e\, Q$ (N3LO) consist of: (i) terms generated by minimal
substitution in the four-nucleon contact interactions involving two gradients of the
nucleon fields as well as by non-minimal couplings to the electromagnetic field;
(ii) OPE terms induced by $\gamma \pi N$ interactions of sub-leading order; and
(iii) one-loop two-pion-exchange (TPE) terms.  We discuss them below.

The contact minimal and non-minimal currents, denoted by the
subscripts \lq\lq min'' and \lq\lq nm'' respectively, are written as~\cite{Piarulli:2013}
\begin{eqnarray}
\label{eq:jmin}
{\bf j}^{(1)}_{\rm min}({\bf q})&=&\frac{i\, e}{16}\, G_E^V(q_\mu^2)\, \left({\bm \tau}_1\times{\bm \tau}_2\right)_z\,
\Big[ (C_2+3\, C_4+C_7)\, {\bf k}_1 \nonumber\\
&&+( C_2-C_4-C_7)\, {\bf k}_1 \,\,{\bm \sigma}_1\cdot{\bm \sigma}_2 \nonumber\\
&& +C_7\, {\bm \sigma}_1\cdot ({\bf k}_1-{\bf k}_2)\,\, {\bm \sigma}_2 \Big]
-\frac{i\, e}{4}\, e_{N,1}(q_\mu^2)\,C_5\, \nonumber\\
 &&\times ({\bm \sigma}_1+{\bm \sigma}_2)
\times {\bf k}_1 + 1 \rightleftharpoons 2 \ , \\
\label{eq:nmcounter}
{\bf j}^{(1)}_{\rm nm}({\bf q})&=& - i\,e \Big[  G_E^S(q_\mu^2)\,C_{15}^\prime\, {\bm \sigma}_1 
+ G_E^V(q_\mu^2)\, C_{16}^\prime\nonumber\\
 &&\times (\tau_{1,z} - \tau_{2,z})\,{\bm \sigma}_1  \Big]\times {\bf q}  
+ 1 \rightleftharpoons 2 \ .
\end{eqnarray}
The low-energy constants (LECs) $C_1,\dots, C_7$ enter the two-nucleon ($2N$) contact
interaction (at NLO), and are constrained by fits to the $np$ and $pp$ elastic scattering
data and the deuteron binding energy.  We take their values from the various $2N$ interactions
we have adopted in the present study (see Sec.~\ref{sec:fit}). We should point out
that in the case of the NV models, the original parametrization
of the contact interaction at NLO is subjected to a Fierz rearrangement, so as
to make it local~\cite{Piarulli:2015}.  As a consequence, the
LECs $C_2$, $C_4$, $C_5$, and $C_7$ in Eq.~(\ref{eq:jmin})
are related to those introduced in Ref.~\cite{Piarulli:2015,Piarulli:2016}
and denoted with a $P$ superscript here for clarity via
\begin{eqnarray}
C_2&=&-4\,C_2^P -12\, C_4^P + 6\, C_6^P\\
C_4&=&-4\,C_2^P+4\,C_4^P+14\, C_6^P \ ,
\end{eqnarray}
$C_5\,$=$\,C_7^P$, and $C_7\,$=$\,-24\,C^P_6$.  In Ref.~\cite{Schiavilla:2019}
this correspondence between
the $C_i$ and $C_i^P$ has been inadvertently ignored.  The
error affects the contribution labeled N3LO(MIN) in Tables~III and IV,
and the fitted values $d_1^S$, $d_2^S$, and $d_1^V$ in Table I of
Ref.~\cite{Schiavilla:2019}.  However, we have verified
that it does not significantly change
the predicted values for the various observables considered
in that work, or alter the ensuing discussions.\footnote{Tables of the corrected values of the
$d_1^S$, $d_2^S$, and $d_1^V$ LECs
are available upon request.}

The LECs $C_{15}^\prime$
and $C_{16}^\prime$ (as well as $d_8^\prime$, $d_9^\prime$, and $d_{21}^\prime$ below) are determined
by fitting measured photo-nuclear observables of the $A=2$ and $3$ systems, as
discussed in Sec.~\ref{sec:fit}.  Finally, there is no {\it a priori} justification for the
use of $G_E^S/G_E^V$ (or $G_M^S/G_M^V$) in the non-minimal contact current;
they are included so as to provide a reasonable fall-off with increasing $q_\mu^2$
for the strength of this current. 

The sub-leading OPE currents at N3LO have isovector (IV) and isoscalar (IS) components
given by, respectively,
\begin{eqnarray}
\label{eq:cdlt}
{\bf j}^{(1)}_{\pi{\rm IV}}({\bf q})&=&i\, e\, \frac{g_A}{4\,f_\pi^2} \frac{G_{\gamma N\Delta}(q_\mu^2)}{\mu_{\gamma N\Delta}}\,
\frac{{\bm \sigma}_2 \cdot {\bf k}_2}{\omega_{k_2}^2}\,
\bigg[  d_8^\prime \,\tau_{2,z} \,{\bf k}_2 \\
&&-d_{21}^\prime\, ({\bm \tau}_1\times{\bm \tau}_2)_z\, {\bm \sigma}_1\times {\bf k}_2  \bigg] \times {\bf q} + 1\rightleftharpoons 2 \ ,
\nonumber
\end{eqnarray}
and
\begin{equation}
{\bf j}^{(1)}_{\pi{\rm IS}}({\bf q})\!=\! i\, e\, \frac{g_A}{4\,f_\pi^2} \, d^\prime_9\,G_{\gamma\pi\rho}(q_\mu^2) \, 
 {\bm \tau}_1\cdot{\bm \tau}_2\,\frac{{\bm \sigma}_2 \cdot {\bf k}_2}{\omega_{k_2}^2}\,
  {\bf k}_2 \times {\bf q}
+ 1\rightleftharpoons 2 \ ,
\label{eq:cp}
\end{equation}
and depend on the three (unknown) LECs $d_8^\prime$, $d_9^\prime$, and $d_{21}^\prime$.
The LECs $d_8^\prime$ and $d_{21}^\prime$ can be related~\cite{Pastore:2009} to the
$N$-$\Delta$ transition axial coupling constant and magnetic moment in a resonance
saturation picture, which justifies the use of the $\gamma N\Delta$ electromagnetic form
factor for this term.  However, we emphasize that, in contrast to
Ref.~\cite{Schiavilla:2019}, $\Delta$ saturation for these LECs is not assumed here.\footnote{In
other words, in the present study, when using the NV interactions,
we include both ${\bf j}^{(0)}_\Delta$ and
${\bf j}^{(1)}_{\pi {\rm IV}}$, whereas in Ref.~\cite{Schiavilla:2019} we only included ${\bf j}^{(0)}_\Delta$.
However, with the interactions of Ref.~\cite{Entem:2017} we only consider ${\bf j}^{(1)}_{\pi {\rm IV}}$. }
The LEC $d^\prime_9$ reduces, in a resonance saturation picture, to the well known
$\gamma\pi\rho$ current~\cite{Pastore:2009}.  Accordingly, we have accounted for
the $q_\mu^2$ fall-off of the electromagnetic vertex by including a $\gamma \pi \rho$
form factor, which in vector-meson dominance is parametrized as
\begin{equation}
G_{\gamma\pi\rho}(q_\mu^2)=\frac{1}{1+q_\mu^2/m_\omega^2} \ ,
\end{equation}
where $m_\omega$ is the $\omega$-meson mass.

The one-loop TPE currents are written as~\cite{Pastore:2009,Piarulli:2013}
\begin{eqnarray}
\label{eq:jloop}
{\bf j}^{(1)}_{2\pi}({\bf q})\!\! &=&\!\! -i\, e\, G_E^V(q_\mu^2)\,\bigg[ ({\bm \tau}_1\times {\bm \tau}_2)_z \,
{\bm \nabla}_{\! k}\, F_1(k) - \tau_{2,z} \\
&&\times  \bigg[ F_0(k) \,{\bm \sigma}_1 
 - F_2(k)\, \frac{{\bf k}\,{\bm \sigma}_1\cdot{\bf k}}{k^2}\bigg]\times {\bf q} \bigg]
+ 1 \rightleftharpoons 2 \ ,
\nonumber
\end{eqnarray}
where ${\bf k}$ is the relative momentum defined above, and the functions $F_i(k)$ are
\begin{eqnarray}
F_0(k)\!&=&\!\frac{g_A^2}{128\,\pi^2 f_{\pi}^4}\Bigg[
1 \!-\!2\, g_A^2+\frac{  8\,g_A^2\, m_\pi^2 }{k^2+4\, m_\pi^2}
+G(k)\bigg[ 2-2\, g_A^2 \nonumber\\
&&-\frac{  4\,(1+g_A^2)\, m_\pi^2 }{k^2+4\, m_\pi^2} 
+\frac{16\, g_A^2 \,m_\pi^4 }{(k^2+4\, m_\pi^2)^2} \bigg] \Bigg]\ ,
\label{eq:f0k} \\
F_1(k)&=&\frac{1}{1536 \, \pi^2\,f_{\pi}^4}\,
G(k) \bigg[4 m_{\pi}^2(1+4 g_A^2-5 g_A^4)\nonumber\\
&&+k^2(1+10 g_A^2 - 23 g_A^4)
-\frac{48\, g_A^4 m^4_\pi}
{4\,  m^2_\pi+k^2}\bigg] \ ,
\label{eq:f1k}\\
F_2(k)&=&\frac{g_A^2}{128\,\pi^2f_{\pi}^4} \Bigg[
2-6\, g_A^2+ \frac{  8\,g_A^2\, m_\pi^2 }{k^2+4\, m_\pi^2}
+G(k) \bigg[4\, g_A^2\nonumber\\
&&-\frac{  4\,(1+3\, g_A^2)\, m_\pi^2 }{k^2+4\, m_\pi^2} 
+\frac{16\, g_A^2 \,m_\pi^4 }{(k^2+4\, m_\pi^2)^2} \bigg] \Bigg]\ , 
\label{eq:f2k}
\end{eqnarray}
with the loop function $G(k)$ defined as
\begin{equation}
G(k)=\frac{\sqrt{4\,m_{\pi}^2+k^2}}{k}\ln 
\frac{\sqrt{4\,m_{\pi}^2+k^2}+k}{\sqrt{4\,m_{\pi}^2+k^2}-k} \ .
\label{eq:loopf}
\end{equation}
As noted in Refs.~\cite{Pastore:2009,Piarulli:2013}, the expression
above follows from expanding the TPE current in powers
of the external field momentum ${\bf q}$ and in retaining up to linear
terms in ${\bf q}$.  It satisfies current conservation with the
TPE $2N$ interaction at NLO.

These currents have power law behavior at large momenta, and
need to be regularized before their matrix elements between nuclear
wave functions can be calculated (incidentally, we note that these
calculations are done here in $r$-space).  We adopt two different regularization schemes
depending on whether the interactions used to generate the wave functions are in
momentum space---the $2N$ N4LO interactions of Refs.~\cite{Entem:2017}---or
in configuration space---the $2N$ N3LO interactions of Ref.~\cite{Piarulli:2016}.
In the former case, the momentum-space two-body currents are multiplied by
a cutoff function of the form $C_\Lambda(p)\,$=$\,{\rm e}^{-(p/\Lambda)^4}$
with $p$ equal to the momentum transfer $k_i$ to nucleon $i$ in the seagull piece
of Eq.~(\ref{eq:nlo1}) and in Eqs.~(\ref{eq:jmin}), and (\ref{eq:cdlt})--(\ref{eq:cp}),
or equal to the relative momentum $k$ in the pion-in-flight term of Eq.~(\ref{eq:nlo1})
and in Eqs.~(\ref{eq:nmcounter}) and~(\ref{eq:jloop}).  As already noted, we do not
retain the $\Delta$-excitation current of Eq.~(\ref{eq:cdlta}) with the interactions
of Ref.~\cite{Entem:2017}. Fourier transforms are then reduced to one-dimensional
integrations, which are easily carried out numerically.

When using the chiral interactions of Ref.~\cite{Piarulli:2016},
we first carry out the Fourier transforms of the various currents, resulting in
configuration-space operators which are highly singular at vanishing inter-nucleon separations, and
then remove this singular behavior by multiplying the various terms by appropriate
$r$-space cutoff functions, identical to those used in Ref.~\cite{Piarulli:2016}
for the $2N$ interaction. The procedure as well as the explicit expressions for the
resulting currents can be found in Ref.~\cite{Schiavilla:2019}.

\section{Determination of low-energy constants}
\label{sec:fit}

In this section we proceed to the determination of the
LECs $d'_8$, $d'_9$, $d'_{21}$, $C'_{15}$ and $C'_{16}$
entering the current operators at N3LO.
Before discussing fitting procedures, we introduce
the various nuclear interaction models and provide
references to the numerical
methods used in the calculation of the various observables.

In this study we consider two different sets of interaction models.
The first consists of the Norfolk models (denoted as NV)
from Ref.~\cite{Piarulli:2016}.  These are N3LO chiral interactions
that include, beyond pion and nucleon, $\Delta$-isobar degrees of freedom explicitly. 
They are formulated in configuration space
with two regulators, $R_{\rm S}$ and $R_{\rm L}$ respectively, for the short-range (contact) component
and long-range (one- and two-pion
exchange) component.  
There are two classes of NV interactions, which differ in the range of lab energy over which
the fits to the $2N$ database were carried out.  The first, denoted as NVI, was fitted in the range
0--125 MeV, while for the second, denoted as NVII, this range was extended to 200 MeV.
Within each class, two different sets of cutoff values $R_{\rm S}$ and $R_{\rm L}$ were considered,
and the resulting models were designated as NVIa/b or NVIIa/b (see Table~\ref{tab:potlist}). 

The second set of interaction models are those developed by
Entem, Machleidt and Nosyk (EMN) in Ref.~\cite{Entem:2017}.  These are
momentum-space chiral interactions including only pions and nucleons as
degrees of freedom. The regularization is implemented in momentum space as well, and
as a consequence these models, in contrast to the NV ones, are strongly non-local in configuration
space.  For this set of interactions, which have all been fitted to the $2N$
database up to 300 MeV lab energy, three different cutoff values, respectively
$\Lambda\,$=$\,450$, $500$ and $550$ MeV, are considered.  All chiral orders 
are available for these models and so, as discussed below, fits of the
electromagnetic LECs were carried out order-by-order from NLO to N4LO.
However, we will report only the values corresponding to the N4LO interactions, if not
otherwise specified.\footnote{Values for the electromagnetic LECs obtained
with the lower order EMN interactions are available from the authors upon request.}
All the interactions considered in the present study along with their main features are listed in Table~\ref{tab:potlist}.
\begin{table}
  \centering
  \begin{tabular}{cccccc}
    \hline
    \hline
    Name & DOF & $O_\chi$ & $(R_{\rm S},R_{\rm L})$ or $\Lambda$ & $E$ range & Space \\
    \hline
    NVIa  & $\pi,N,\Delta$ & N3LO & $(0.8,1.2)$ fm & 0--125 MeV & $r$\\
    NVIb  & $\pi,N,\Delta$ & N3LO & $(0.7,1.0)$ fm & 0--125 MeV & $r$\\
    \hline
    NVIIa & $\pi,N,\Delta$ & N3LO & $(0.8,1.2)$ fm & 0--200 MeV & $r$\\
    NVIIb & $\pi,N,\Delta$ & N3LO & $(0.7,1.0)$ fm & 0--200 MeV & $r$\\
    \hline\hline
    EMN450     & $\pi,N$        & N4LO & $450$ MeV      & 0--300 MeV & $p$\\
    EMN500     & $\pi,N$        & N4LO & $500$ MeV      & 0--300 MeV & $p$\\
    EMN550     & $\pi,N$        & N4LO & $550$ MeV      & 0--300 MeV & $p$\\
    \hline
    \hline
  \end{tabular}
  \caption{\label{tab:potlist}Summary of $2N$ interaction models utilized in this
study. In the first column we indicate the name adopted to identify each model
and in the remaining columns its main features, including degrees of freedom (DOF), chiral order ($O_\chi$),
cutoff values, lab-energy range over which the fits to the $2N$ database have been carried out ($E$ range), and
whether it is in configuration ($r$) or in momentum ($p$) space. }
\end{table}

In the calculations of $A\,$=$\,3$ observables, we include (chiral) three-nucleon ($3N$) interactions at
LO (in particular, for the NV models we also account for the LO two-pion-exchange term originating from
$\Delta$-resonance intermediate states~\cite{Piarulli:2018}).  The LECs ($c_D$ and $c_E$, in standard notation)
that characterize them have been constrained by reproducing the $\Ht$-$\Het$ binding energies and the
Gamow-Teller matrix element in tritium beta decay.  We take the LECs from Table III of Ref.~\cite{Baroni:2018}
for the NV set of $2N$ interactions and from Table II of Ref.~\cite{Marcucci:2019} for the EMN set.
In the following, we will denote with the $^*$ superscript the NV Hamiltonians consisting of these $2N$ and $3N$
interactions.  The $^*$ is to differentiate them from the NV Hamiltonians in which the
$3N$ interactions are fitted to reproduce a different set of trinucleon observables~\cite{Piarulli:2018}
($^3$H-$^3$He binding energies and $nd$ doublet scattering length).
The $A\,$=$\,3$ wave functions have been obtained from solutions of the Schr\"odinger equation with
the hyperspherical-harmonic approach~\cite{Kievsky:2008,Marcucci:2020}.

The calculation of the $A\,$=$\,2$--3 magnetic form factors and magnetic moments is
performed using Monte Carlo integration techniques~\cite{Schiavilla:1989}.  The number
of sampled configurations utilized is of the order of $10^6$ for the deuteron and
$5\times10^5$ for the $A\,$=$\,3$ systems.  The statistical errors due to these
Monte Carlo integrations are typically $\lesssim 1\%$ over the range of momentum transfers considered.
In the final results these errors are summed up in quadrature with uncertainties from other
sources we account for in the present study (see below).

The  cross section for the deuteron threshold
electrodisintegration at backward angles ($d$-threshold) has been obtained
by evaluating the relevant matrix elements of the electromagnetic current
between the ground state and $np$ continuum states using standard Gaussian
quadrature methods~\cite{Schiavilla:2019,Schiavilla:1991}.
While the $d$-threshold experimental data have been averaged over the interval 0--3 MeV
of the final $np$-pair center-of-mass energy, the theoretical
results have been computed at a fixed energy of 1.5 MeV.  It
is known that the effect of the width of the energy bin
over which the cross-section values are averaged is
small~\cite{Schiavilla:1991}.

We can now discuss the fit of the electromagnetic LECs entering the OPE
and non-minimal contact currents at N3LO. We introduce the set $d^{S}_i$ and $d^{V}_i$
of adimensional LECs---the superscript specifies the isoscalar ($S$) or isovector
($V$) character of the associated operators---in units of $m_\pi$ (cutoff
$\Lambda$) for the NV (EMN) Hamiltonians, namely
\begin{equation}
\begin{aligned}
&C_{15}^\prime=d_1^S/\lambda^4  \ , \quad d_9^\prime=d_2^S/\lambda^2 \\
&C_{16}^\prime=d_1^V/\lambda^4 \ , \quad d_8^\prime=d_2^V/\lambda^2\ , \quad d_{21}^\prime=d_3^V/\lambda^2 \ ,
\end{aligned}
\end{equation}
with $\lambda\,$=$\,m_\pi$ or $\Lambda$ for
the NV or EMN Hamiltonians, respectively.
Their corresponding values are reported in Tables~\ref{tab:fitNV} and~\ref{tab:fitEM}. We fitted the LECs using a $\chi^2$ minimization procedure on two different sets of data.   Set A includes the
magnetic moments of deuteron ($\mu_d$), tritium ($\mu_{\Ht}$), helium ($\mu_{\Het}$)~\cite{NNDC} and the $d$-threshold 
cross sections data up to $q_\mu^2\,$=$\,40$ fm$^{-2}$~\cite{Rand:1967,Ganichot:1972,Bernheim:1981,Auffret:1985,Schmitt:1997}, whereas set B includes
the A data set plus the deuteron magnetic form factor $G_M(q_\mu)$ data up to momentum transfers of
$q_\mu\,$=$\,6$ fm$^{-1}$.  The star superscript indicates that from
the sets A and B we have removed
the $d$-threshold cross-section data corresponding to the Rand {\it et al.}~experiment of 1967~\cite{Rand:1967}.
While these data do not impact significantly the determination of
the LECs, they do produce a large increase in the $\chi^2$/datum.

In the present study we carry out a simultaneous fit of these data sets,
without separating them in (predominantly) isoscalar and isovector observables.
Thus, we are able to account explicitly for the fact that the trinucleons are not pure isospin $T\,$=$\,1/2$ states,
but also include small $T\,$=$\,3/2$ admixtures, induced by isospin symmetry-breaking interaction
terms present in the NV and EMN Hamiltonians.  Furthermore, in the $d$-threshold
cross-section calculations, we include all continuum states and not just the dominant
$^1$S$_0$ channel.  Note that the present strategy for constraining the LECs is different
from that adopted in Refs.~\cite{Piarulli:2013,Schiavilla:2019}, where resonance saturation
arguments were invoked to reduce their number and only the magnetic moments of $A\,$=2
and 3 nuclei and/or the radiative capture cross section for thermal neutrons on protons were fitted.

The LECs of Tables~\ref{tab:fitNV} and~\ref{tab:fitEM} are generally in line with the
expected values based on the naive power counting given in Ref.~\cite{Schiavilla:2019}.
However, $d_2^V$ is too large by a factor of 10--20 for all Hamiltonian
models we have considered, as is $d_2^S$ corresponding to the data sets A and A$^*$
for the EMN500 and EMN550 models.  Lastly, the LEC $d_1^V$ for the EMN500 and EMN550 interactions
  is poorly constrained by the fits, and so it is difficult
to have a meaningful comparison with the naive power counting.
\begin{table}
  \begin{center}
    \begin{tabular}{lcccccc}
      \hline\hline
         Set & $d_1^V$ & $d_2^V$ & $d_3^V$ & $d_1^S$  & $d_2^S$ & $\chi^2/N$\\ 
      \hline
      \multicolumn{7}{c}{NVIa$^*$}\\
      \hline
        A           & $-0.050(2)$ & $\m0.49(7)$ & $0.094(4)$ & $0.012(1)$ & $0.023(7)$ &$\z11.0$ \\
        A$^\star$   & $-0.050(2)$ & $\m0.59(6)$ & $0.109(4)$ & $0.012(1)$ & $0.028(9)$ &$\z2.6$  \\
        B           & $-0.052(1)$ & $\m0.45(3)$ & $0.093(3)$ & $0.011(1)$ & $0.032(8)$ &$12.4$ \\
        B$^\star$   & $-0.050(1)$ & $\m0.61(6)$ & $0.114(6)$ & $0.009(1)$ & $0.044(6)$ &$\z3.9$ \\
      \hline
      \multicolumn{7}{c}{NVIb$^*$}\\
      \hline
        A          & $-0.055(3)$ & $\m0.09(5)$ & $0.073(3)$ & $0.025(2)$ & $ 0.030(6)$ &$11.0$ \\
        A$^\star$  & $-0.053(1)$ & $\m0.18(3)$ & $0.086(3)$ & $0.029(2)$ & $ 0.044(8)$  &$\z2.7$  \\
        B          & $-0.057(1)$ & $\m0.07(3)$ & $0.073(3)$ & $0.026(2)$ & $0.038(6)$  &$12.0$ \\
        B$^\star$  & $-0.052(2)$ & $\m0.21(4)$ & $0.090(4)$ & $0.030(2)$ & $0.052(5)$  &$\z3.6$ \\
      \hline
      \multicolumn{7}{c}{NVIIa$^*$}\\
      \hline
        A         & $-0.066(2)$ & $\m0.01(7)$ & $0.069(4)$ & $0.011(1)$ & $0.019(7)$ &$11.0$ \\
        A$^\star$ & $-0.064(3)$ & $\m0.15(9)$ & $0.086(5)$ & $0.011(1)$ & $0.020(8)$ &$\z2.6$  \\
        B         & $-0.068(2)$ & $ -0.03(5)$ & $0.069(3)$ & $0.010(1)$ & $0.025(8)$ &$12.0$ \\
        B$^\star$ & $-0.067(1)$ & $\m0.10(5)$ & $0.087(5)$ & $0.009(1)$ & $0.040(8)$ &$\z3.5.$ \\
      \hline
      \multicolumn{7}{c}{NVIIb$^*$}\\
      \hline
        A         & $-0.049(2)$ & $\m0.09(4)$ & $0.048(3)$ & $0.017(1)$ & $0.018(3)$ &$12.5$ \\
        A$^\star$ & $-0.048(3)$ & $\m0.17(5)$ & $0.060(4)$ & $0.019(2)$ & $0.022(4)$ &$\z3.2$  \\
        B         & $-0.050(3)$ & $\m0.08(4)$ & $0.047(3)$ & $0.017(1)$ & $0.020(3)$ &$13.4$ \\
        B$^\star$ & $-0.050(3)$ & $\m0.14(3)$ & $0.058(3)$ & $0.020(2)$ & $0.027(5)$ &$\z4.0.$ \\
      \hline\hline
    \end{tabular}
    \caption{\label{tab:fitNV} Adimensional values of the
      LECs corresponding to the NV Hamiltonians Ia$^*$, Ib$^*$, IIa$^*$, and IIb$^*$ fitted to various data
      sets; see text for more details.}
  \end{center}
\end{table}
\begin{table}
  \begin{center}
    \begin{tabular}{lcccccc}
      \hline\hline
         Set & $d_1^V$ & $d_2^V$ & $d_3^V$ & $d_1^S$  & $d_2^S$ & $\chi^2/N$\\ 
      \hline
      \multicolumn{7}{c}{EMN450}\\
      \hline
       A           & $\m1.9(2)$ & $6.2(1)$ & $0.39(2)$ & $2.4(3)$ & $\m0.08(10)$ &$11.3$ \\
       A$^\star$   & $\m2.9(2)$ & $6.7(1)$ & $0.51(2)$ & $2.0(3)$ & $\m0.23(9) $ &$\z2.8$\\
       B           & $\m2.4(2)$ & $6.4(1)$ & $0.43(2)$ & $3.32(6)$& $ -0.29(3) $ &$13.0$ \\
       B$^\star$   & $\m3.8(2)$ & $7.2(1)$ & $0.60(2)$ & $3.36(6)$& $ -0.31(3) $ &$\z4.4$ \\
      \hline
      \multicolumn{7}{c}{EMN500}\\
      \hline
       A           & $ -1.2(6)$ & $4.3(3)$ & $0.55(3)$ & $2.2(3)$  & $\m1.5(2)$ &$14.7$ \\
       A$^\star$   & $ -0.5(6)$ & $4.6(3)$ & $0.65(3)$ & $2.2(2)$  & $\m1.5(1)$ &$\z4.7$  \\
       B           & $ -0.2(6)$ & $4.9(3)$ & $0.57(3)$ & $2.08(4)$ & $\m0.33(7)$&$34.3$ \\
       B$^\star$   & $\m0.5(6)$ & $5.2(3)$ & $0.67(3)$ & $2.09(4)$ & $\m0.35(7)$&$24.5$ \\
      \hline
      \multicolumn{7}{c}{EMN550}\\
      \hline
       A          & $ -0.6(1.5)$ & $5.7(6)$ & $0.31(5)$ & $5.2(2)$ & $\m6.2(4)$ &$17.7$ \\
       A$^\star$  & $\m0.2(1.4)$ & $6.0(6)$ & $0.41(5)$ & $5.3(2)$ & $\m6.4(4)$ &$\z7.9$  \\
       B          & $ -0.3(1.5)$ & $5.8(6)$ & $0.26(5)$ & $2.4(1)$ & $-0.6(1)$  &$34.1$ \\
       B$^\star$  & $\m0.8(1.5)$ & $6.3(6)$ & $0.36(5)$ & $2.4(1)$ & $-0.6(1)$  &$24.9$ \\
      \hline\hline
    \end{tabular}
    \caption{\label{tab:fitEM} Same as for Table~\ref{tab:fitNV}, but for 
      the EMN Hamiltonians corresponding to
      $\Lambda\,$=$\,450,500$, and $550$ MeV.}
  \end{center}
\end{table}

\begin{figure}[bth]
\includegraphics[scale=0.65]{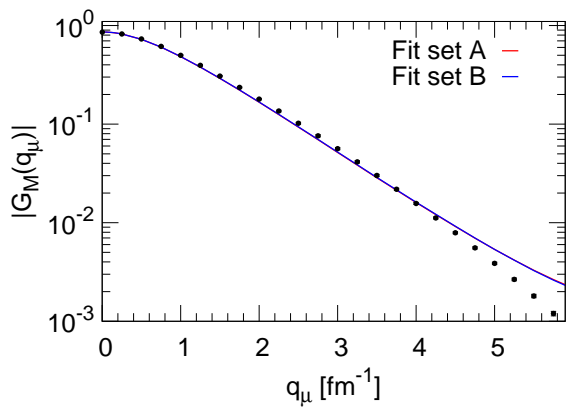}
\caption{\label{fig:deutcomp}Deuteron magnetic form factor computed using the NVIa interaction
  and corresponding to set A or B of LECs.  The difference between
  the two curves cannot be appreciated. Similar results
  have been obtained with all the other interactions we consider. For clarity, errors
  have not been included in the figure.}
\end{figure}

In an attempt to improve the description of the deuteron $G_M(q_\mu)$ data
at higher momentum transfers (thus, ``stretching'' significantly the regime of applicability
of the present $\chi$EFT framework), we included these data in the fits of the LECs (set B).
However, this has a completely negligible impact, as can be seen in
Fig.~\ref{fig:deutcomp} where we compare the results for $G_M(q_\mu)$
obtained using the LECs corresponding to sets A and  B.  We also note that including
the deuteron $G_M(q_\mu)$ data in the fits does not alter the description
of all remaining observables we consider in the present study.

By removing the $d$-threshold data of Ref~\cite{Rand:1967} (sets A$^*$ and B$^*$ of LECs),
the reduced $\chi^2$ decreases very substantially, reaching a value of $\approx 3$ for the NV
interactions (the total number of data is of the order of 40 for set A and 65 for set B).
It is worthwhile remarking here that we are fitting
cross sections that fall by several orders of magnitude and
span a region of momentum transfers $q_\mu^2$ extending up to 40 fm$^{-2}$,
in fact, well beyond the regime of validity of $\chi$EFT.  We also note that, overall,
the quality of the fits for the other observables as well as the predictions for the $A\,$=$\,2$--3
nuclei magnetic form factors do not differ appreciably when using the un-starred
and starred sets of LECs (the only exceptions are the EMN550-based predictions
for the trinucleon magnetic form factors, see below).
For this reason, hereafter
we will discuss only the results obtained with set A of LECs.

Individual contributions, associated with the various terms
of the current, to $\mu_d$ , $\mu_{\Ht}$, and $\mu_{\Het}$ ,
are reported in Tables~\ref{tab:mud},~\ref{tab:mutri} and~\ref{tab:mu3he}, respectively.
The quoted errors are obtained by propagating the uncertainties on the fitted LECs.
The Monte Carlo statistical errors are also included (in quadrature) in the uncertainties quoted for
the total results.  The theoretical errors of these observables are less than one \%.
In general, the fit is able to reproduce a value compatible with experiment within the theoretical error bars,
except for the EMN500 and EMN550 interactions in the
deuteron magnetic moment case.

The $\mu_d$ observable receives contributions only from the isoscalar terms of the
current.  The results corresponding to the NV interactions obtained for the
LO and N2LO are identical to those given in Ref.~\cite{Schiavilla:2019}, but not for the
N3LO(min) contribution or the contributions
proportional to the refitted LECs $d^S_1$ and $d_2^S$, for the reason explained above.
Inspection of Table~\ref{tab:mud}
shows that (for the NV interactions) the N3LO OPE correction of Eq.~(\ref{eq:cp})
has opposite sign for models a and b, as in Ref.~\cite{Schiavilla:2019}, where
the origin of this sign flip is explained (see Fig.~3 of that paper).  A sign flip also results
between EMN450 or EMN500 and EMN550.  Generally, 
the $\mu_d$'s obtained with the EMN interactions tend to overestimate the experimental value.
The main reason seems to be that the fits based on these models are not able to constrain
the LEC $d_2^S$ as for the NV cases.

Next, we discuss the results for the $A\,$=$\,3$ magnetic moments in Tables~\ref{tab:mutri}
and~\ref{tab:mu3he}.  We note that the row labeled N2LO includes only the contribution of
the relativistic single-nucleon current (${\bf j}^{(0)}_{\rm RC}$ in the notation of the previous section)
for the EMN models.  However, in the case of the NV models, this row also includes
the contribution of the $\Delta$-excitation current (${\bf j}_\Delta^{(0)}$),
which is dominant at N2LO and responsible for the sign flip.
Both the NV and EMN interactions yield excellent fits of
the experimental values of the trinucleon magnetic
moments.
\begin{table*}
  \begin{center}
    \begin{tabular}{llllllll}
      \hline\hline
      & NVIa$^*$ & NVIb$^*$ & NVIIa$^*$ & NVIIb$^*$  & EMN450 & EMN500 & EMN550 \\
      \hline                                                                                                                                        
      LO            & $\m0.8499    $ & $\m0.8486    $ & $\m0.8500   $ & $\m0.8501  $ & $\m0.8549   $ & $\m0.8564   $ & $\m0.8562    $ \\
      N2LO          & $ -0.0062    $ & $ -0.0062    $ & $ -0.0065   $ & $ -0.0071  $ & $ -0.0069   $ & $ -0.0070   $ & $ -0.0074    $ \\
      N3LO(min)     & $\m0.0284    $ & $\m0.0301    $ & $\m0.0271   $ & $\m0.0242  $ & $\m0.0425   $ & $\m0.0317   $ & $\m0.0330    $ \\
      N3LO($d_1^S$) & $ -0.0115(9) $ & $ -0.021(2)$  & $ -0.0100(9)$ & $ -0.014(1)$ & $ -0.029(4) $ & $ -0.012(2) $ & $ -0.0199(6) $ \\
      N3LO($d_2^S$) & $ -0.0015(4) $ & $\m0.008(2)$  & $ -0.0011(4)$ & $\m0.006(1)$ & $ -0.003(3) $ & $ -0.0049(8)$ & $ \m0.0094(6)$ \\
      Total         & $\m0.859(6)$   & $\m0.860(5)$   & $\m0.860(6)$  & $\m0.860(2)$ & $\m0.859(10)$ & $\m0.864(2) $ & $\m0.871(2)$ \\ 
      Exp.          &\multicolumn{7}{c}{0.8574}\\                                          
      \hline
      \hline
    \end{tabular}                                                                          
    \caption{\label{tab:mud} Individual contributions to the deuteron magnetic
      moment (in units of n.m.) corresponding to the NV and EMN $2N$ interactions. The
      errors on the $d^{S}_i$ terms are only generated by the LECs (set A). The uncertainties quoted
      for the total are given by the sum in quadrature of the uncertainties due to the LECs and
      Monte Carlo integration.}
  \end{center}
\end{table*}
\begin{table*}
  \begin{center}
    \begin{tabular}{llllllll}
      \hline\hline
      & NVIa$^*$ & NVIb$^*$ & NVIIa$^*$ & NVIIb$^*$  & EMN450 & EMN500 & EMN550 \\
      \hline                                                                                                                                        
      LO            &$ \m2.593    $&$ \m2.585    $&$ \m2.592    $&$ \m2.590    $&$ \m2.599   $&$ \m2.620   $&$ \m2.614   $\\
      NLO           &$ \m0.196    $&$ \m0.223    $&$ \m0.195    $&$ \m0.223    $&$ \m0.173   $&$ \m0.213   $&$ \m0.223   $\\
      N2LO          &$ \m0.032    $&$ \m0.057    $&$ \m0.031    $&$ \m0.054    $&$  -0.024   $&$  -0.026   $&$  -0.027   $\\
      N3LO(TPE)     &$ \m0.026    $&$ \m0.018    $&$ \m0.026    $&$ \m0.015    $&$ \m0.064   $&$ \m0.047   $&$ \m0.044   $\\
      N3LO(min)     &$ \m0.041    $&$ \m0.043    $&$ \m0.038    $&$ \m0.035    $&$ \m0.045   $&$ \m0.033   $&$ \m0.035   $\\
      N3LO($d_1^V$) &$ \m0.110(5) $&$ \m0.106(5) $&$ \m0.144(5) $&$ \m0.090(4) $&$  -0.046(4)$&$ \m0.015(8)$&$ \m0.01(1)$\\
      N3LO($d_2^V$) &$ \m0.046(6) $&$ \m0.130(7) $&$ \m0.001(7) $&$ \m0.139(7)$&$ \m0.185(3)$&$ \m0.090(7)$&$ \m0.10(1)$\\
      N3LO($d_3^V$) &$  -0.048(2) $&$  -0.051(2) $&$  -0.036(2) $&$  -0.034(3) $&$ \m0.024(1)$&$ \m0.015(1)$&$ \m0.006(1)$\\
      N3LO($d_1^S$) &$  -0.014(1) $&$  -0.026(2) $&$  -0.013(1) $&$  -0.017(1) $&$  -0.037(5)$&$  -0.015(2)$&$  -0.024(1)$\\
      N3LO($d_2^S$) &$  -0.003(1) $&$ \m0.008(2) $&$  -0.003(1) $&$ \m0.006(1) $&$  -0.004(5)$&$  -0.015(2)$&$  -0.005(0)$\\
      Total         &$ \m2.98(1)  $&$ \m2.98(1)  $&$ \m2.98(1)  $&$ \m2.98(1)  $&$ \m2.98(2) $&$ \m2.97(1)$&$ \m2.97(2)$\\
      \hline
      Exp.      &\multicolumn{7}{c}{2.979}\\
      \hline
      \hline
    \end{tabular}                                                                          
    \caption{\label{tab:mutri} Same as Table~\ref{tab:mud} for the tritium magnetic moment.  Note the row labeled N2LO includes the contributions of both ${\bf j}_{\rm RC}^{(0)}$
    and ${\bf j}_{\Delta}^{(0)}$ for the NV models, but only those of ${\bf j}_{\rm RC}^{(0)}$ for the EMN models.}
  \end{center}
\end{table*}
\begin{table*}
  \begin{center}
    \begin{tabular}{llllllll}
      \hline\hline
      & NVIa$^*$ & NVIb$^*$ & NVIIa$^*$ & NVIIb$^*$  & EMN450 & EMN500 & EMN550 \\
      \hline                                                                                                                                        
      LO              &$ -1.775   $&$  -1.770   $&$  -1.774   $&$  -1.772   $&$  -1.767   $&$  -1.786   $&$  -1.783   $\\
      NLO             &$ -0.193   $&$  -0.221   $&$  -0.193   $&$  -0.220   $&$  -0.171   $&$  -0.211   $&$  -0.220   $\\
      N2LO            &$ -0.044   $&$  -0.070   $&$  -0.044   $&$  -0.068   $&$ \m0.009   $&$ \m0.011   $&$ \m0.011   $\\
      N3LO(TPE)       &$ -0.026   $&$  -0.017   $&$  -0.026   $&$  -0.014   $&$  -0.062   $&$  -0.045   $&$  -0.043   $\\
      N3LO(min)       &$\m0.030   $&$ \m0.032   $&$ \m0.029   $&$ \m0.025   $&$ \m0.062   $&$ \m0.047   $&$ \m0.045   $\\
      N3LO($d_1^V$)   &$ -0.107(4)$&$  -0.104(5)$&$  -0.141(5)$&$  -0.088(4)$&$ \m0.045(4)$&$  -0.014(8)$&$  -0.01(1)$\\
      N3LO($d_2^V$)   &$ -0.045(6)$&$  -0.013(7)$&$  -0.001(6)$&$  -0.014(7)$&$  -0.181(3)$&$  -0.088(7)$&$  -0.10(1)$\\
      N3LO($d_3^V$)   &$\m0.047(2)$&$ \m0.051(2)$&$ \m0.035(2)$&$ \m0.033(2)$&$  -0.023(1)$&$  -0.014(1)$&$  -0.005(1)$\\
      N3LO($d_1^S$)   &$ -0.014(1)$&$  -0.025(2)$&$  -0.012(1)$&$  -0.017(1)$&$  -0.036(5)$&$  -0.015(2)$&$  -0.024(1)$\\
      N3LO($d_2^S$)   &$ -0.003(1)$&$ \m0.008(2)$&$  -0.002(1)$&$ \m0.006(1)$&$  -0.004(5)$&$  -0.014(2)$&$  -0.004(0)$\\
      Total           &$ -2.13(1) $&$  -2.13(1) $&$  -2.13(1) $&$  -2.13(1) $&$  -2.13(2) $&$  -2.13(1) $&$  -2.13(2)$\\
      \hline
      Exp.      &\multicolumn{7}{c}{-2.126}\\
      \hline
      \hline
    \end{tabular}                                                                          
    \caption{\label{tab:mu3he} Same as Table~\ref{tab:mutri} for the helium magnetic moment.}
  \end{center}
\end{table*}

\begin{figure*}[bth]
  \hspace*{-0.5cm}
\includegraphics[scale=0.49]{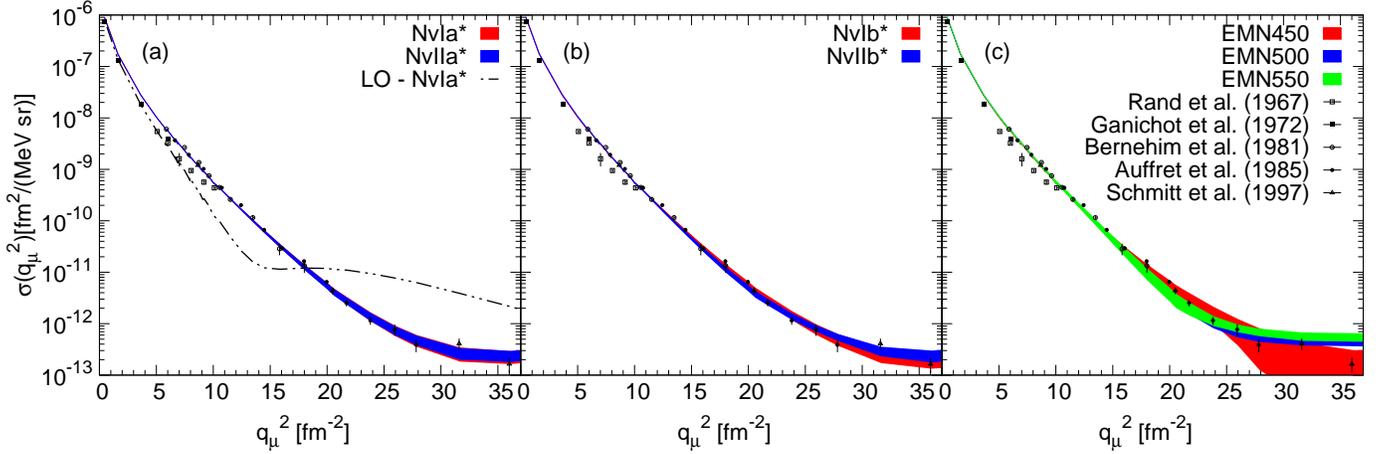}
\caption{\label{fig:dth}The deuteron threshold electrodisintegration data at backward angles are compared to fits corresponding to 
  interaction models NVIa$^*$ and NVIIa$^*$ in panel a, NVIb$^*$ and NVIIb$^*$ in panel b, and EMN450, EMN500, EMN550 in panel c.
  The bands represent the error that has been calculated propagating the uncertainties on the LECs with standard methods.
  The dashed line in panel a is the calculation performed with the LO current for the NVIa$^*$ interaction. }
\end{figure*}

In Fig.~\ref{fig:dth} we present the results of the fit of the $d$-threshold data.
The bands represent the error that has been calculated by propagating the
uncertainties on the LECs with standard methods.  All interactions
reproduce nicely the data up to $q_\mu^2\,$=$\,30$ fm$^{-2}$ (panels a and b for the NV models,
and panel c for the EMN models).
The $q_\mu^2$ behavior is sensitive to the interference between the fitted N3LO terms and the LO
and (remaining) higher order terms in the current.  For both the NV and EMN interactions
the dominant interference is in the isovector
sector for $q_\mu^2$ in the range $10 \lesssim  q_\mu^2 \lesssim 20$ fm$^{-2}$.
For $q_\mu^2\gtrsim20$ fm$^{-2}$, the interference in the isoscalar sector becomes significant and,
in some instances, of the same magnitude as in the isovector sector.  The EMN450 model stands out in
that the interference in the isoscalar sector is dominant.  For this model the fit appears to be unable
to constrain the LEC $d_2^S$, generating the big error band that can be seen in panel c of Figure~\ref{fig:dth}. 

\section{Magnetic form factors of the deuteron and trinucleons}
\label{sec:res}
\begin{figure*}[bth]
  \hspace*{-0.5cm}
\includegraphics[scale=0.49]{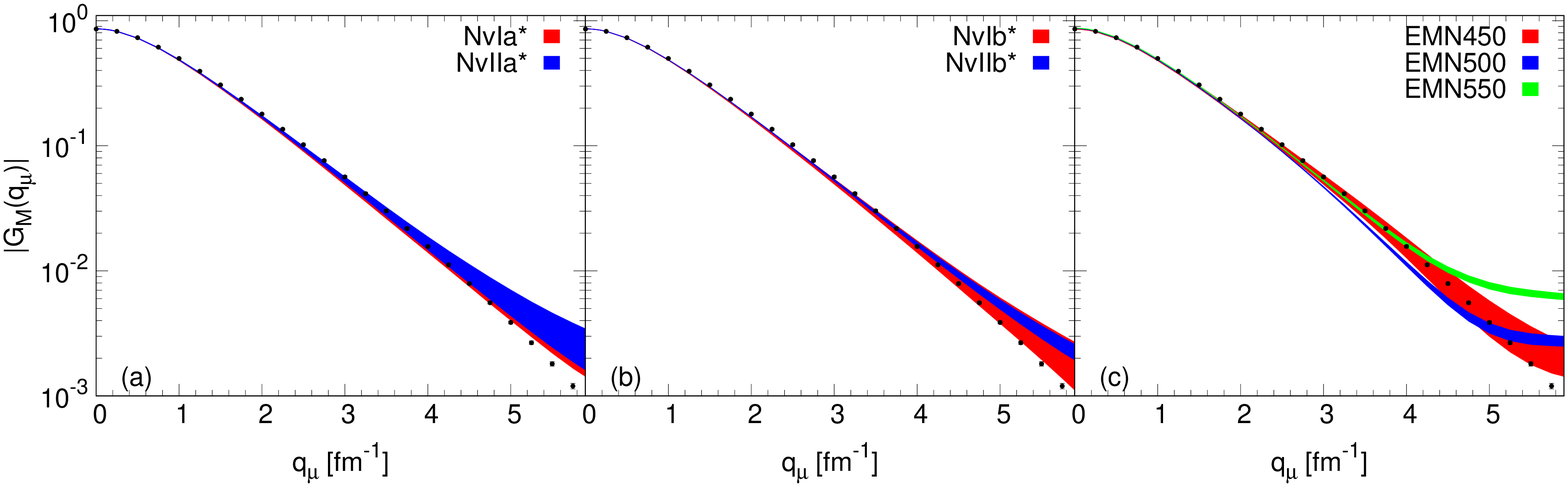}
\caption{\label{fig:FF2}The deuteron magnetic form factor is compared to predictions obtained with
  interaction models NVIa$^*$ and NVIIa$^*$ in panel a, NVIb$^*$ and NVIIb$^*$ in panel b, and EMN450, EMN500, EMN550 in panel c.
  The bands represent the error that has been calculated propagating the uncertainties on the LECs with standard methods.
  Note that in panel a the error bands almost coincide.}
\end{figure*}

\begin{figure*}[bth]
  \hspace*{-0.5cm}
\includegraphics[scale=0.49]{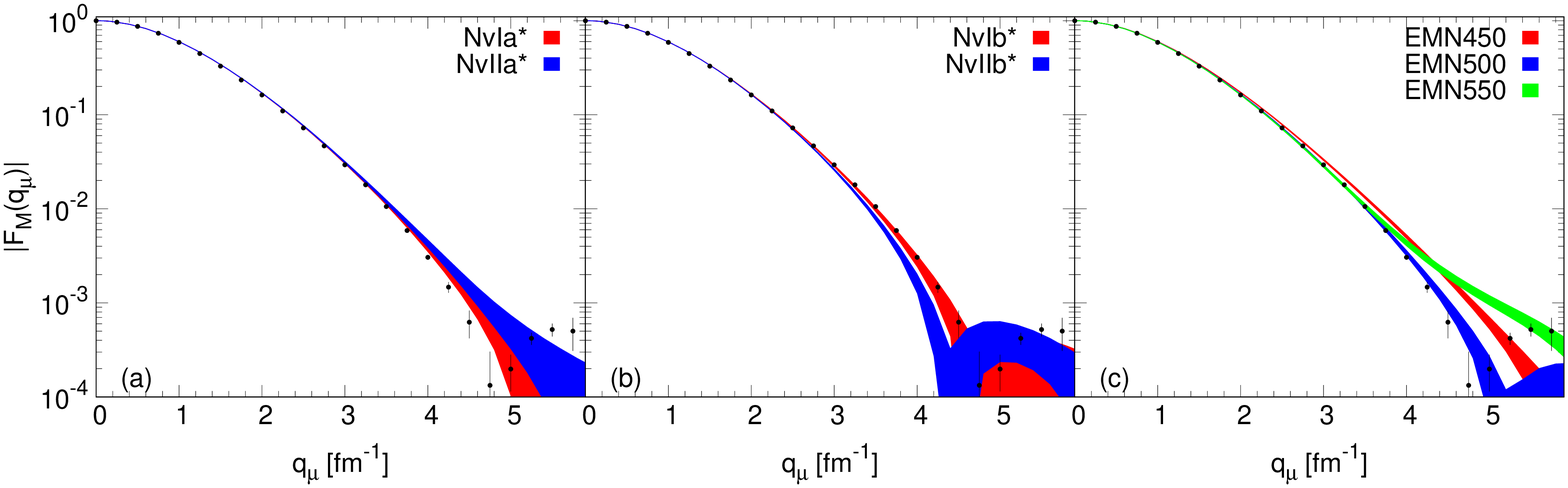}
\caption{\label{fig:FF3H}The same as Figure~\ref{fig:FF2} for the $\Ht$ magnetic form factor.}
\end{figure*}

\begin{figure*}[bth]
  \hspace*{-0.5cm}
\includegraphics[scale=0.49]{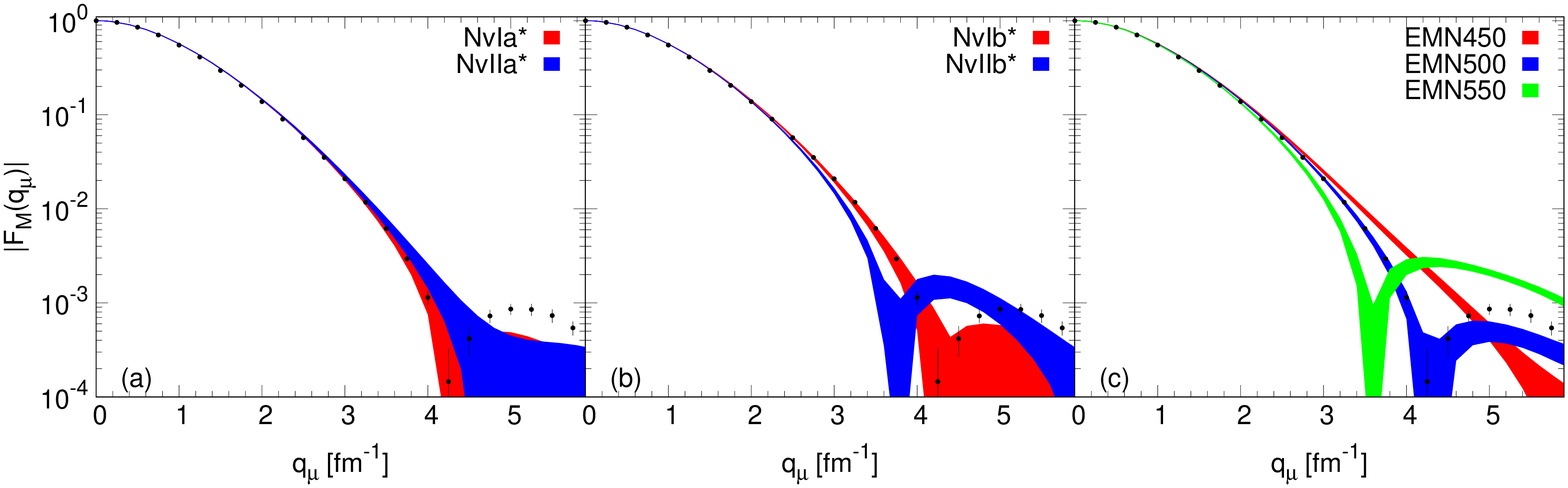}
\caption{\label{fig:FF3He}The same as Figure~\ref{fig:FF2} for the $\Het$ magnetic form factor.}
\end{figure*}

The $d$, $\Ht$ and $\Het$ magnetic form factors corresponding to set A of LECs for all interaction models
used in this study are compared with experimental data in Figs.~\ref{fig:FF2},~\ref{fig:FF3H}
and~\ref{fig:FF3He}, respectively.
In all figures the bands represent the uncertainties coming from the statistical errors in the Monte Carlo
integration (albeit these errors are essentially negligible), and the propagation of the fitted LECs errors
summed in quadrature. The error band expands because the relative weight of the
N3LO terms and associated uncertainties becomes larger and larger as $q_\mu$ increases. 
These magnetic form factors represent predictions of
our models. The trinucleon magnetic form factors have been normalized to unity at $q_\mu\,$=$\,0$.

The deuteron magnetic form factor obtained with the NV interactions is shown
in Figs.~\ref{fig:FF2}a and~\ref{fig:FF2}b. 
Despite the more sophisticated fitting procedure adopted here
than in previous works, we have very good agreement with experimental
data only for $q_\mu\lesssim 3.5$ fm$^{-1}$. At higher $q_\mu$, theory overestimates the data;
  this overestimate is primarily due to the isoscalar spin-orbit term in the min current of Eq.~(\ref{eq:jmin}) that is
  only partially corrected by the fitted (isoscalar) non-minimal term of Eq.~(\ref{eq:nmcounter}).
The isoscalar OPE contribution (at N3LO) is instead almost negligible.

The results for the EMN models in Fig.~\ref{fig:FF2}c show a similar, albeit more pronounced, behavior.
In particular, for EMN500 and EMN550 the form factors
are almost flat for $q_\mu\gtrsim 5$ fm$^{-1}$.  In these cases, the main contribution
is again given by the contact (isoscalar) spin-orbit term proportional to  $C_5$, which is especially large in the
EMN models.  Moreover, the $q_\mu$ falloff comes primarily from that
of $G_E^S(q_\mu^2)$.  The EMN500 also tends to underestimate
the data in the region of $2 \lesssim  q_\mu \lesssim 5$ fm$^{-1}$. 

The NV predictions for the trinucleon magnetic form factors with the present
fitting procedure of LECs are generally in excellent agreement with the data up to
$q_\mu\lesssim4$ fm$^{-1}$ as can be seen in Figs.~\ref{fig:FF3H}a
and~\ref{fig:FF3H}b for $\Ht$ and Figs.~\ref{fig:FF3He}a and~\ref{fig:FF3He}b
for $\Het$  (with the exception of the NVIb$^*$ model). The fitted currents at N3LO are able
to fill in the diffraction region generated at LO (by contrast, see Ref.~\cite{Schiavilla:2019}) and, particularly
in the case of the NVIa$^*$/IIa$^*$ models, reproduce well the minima seen
in the $\Het$ magnetic form factor. 
The interplay between the terms proportional
to the LECs $d_1^V$ and $d_3^V$ and, in particular, the fine tuning between these LECs
play a major role in achieving, at these high $q_\mu$, such a level of success. 
Thus, the new procedure appears to validate the $\chi$EFT modeling of the
electromagnetic current well beyond the $q_\mu \lesssim2$ fm$^{-1}$ limit of
Ref.~\cite{Schiavilla:2019}.  

The picture for the EMN models is somewhat less satisfactory.
The EMN450 model, as in the $d$ case, completely
fails to describe the data much beyond $q\gtrsim3$ fm$^{-1}$.  The EMN500
model provides the best description of the measured form factors, in fact
at the same level of the NV interactions.  The EMN550 reproduces poorly the data
because of the large contribution proportional to $d_2^S$, which becomes
dominant for $q_\mu \gtrsim 3$ fm$^{-1}$. As a matter
of fact, the $d_2^S$ contribution resulting from the fit of the data set B (or B$^*$) is much reduced, and
provides a description of the trinucelon magnetic form factors similar to
that obtained with the EMN500 model.

\begin{figure*}[bth]
\includegraphics[scale=0.6]{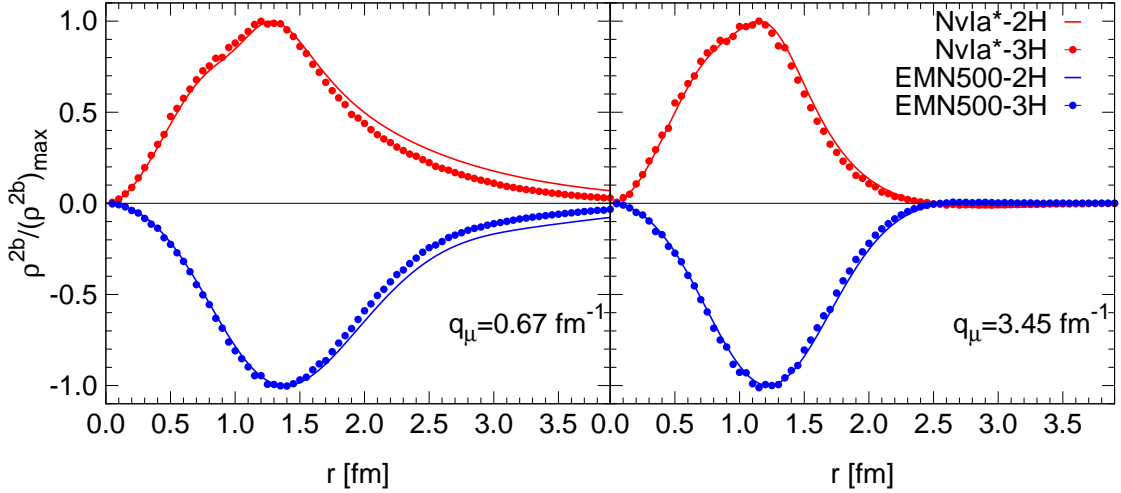}
\caption{\label{fig:density} Two-body transition densities corresponding
to the NVIa$^*$ and EMN500 interaction models.  All curves have been
rescaled so as peak at 1. We inverted the sign of the EMN500 results to
avoid clutter.  The densities are computed at $q_\mu=0.67$ fm$^{-1}$ (left panel) and $q_\mu=3.45$ fm$^{-1}$ (right panel).}
\end{figure*}

Because of the constraint on the LECs provided
by the $d$-threshold cross-section data at high $q_\mu^2$, the
present fitting procedure is much more successful in describing
the magnetic form factors of the trinucleons  at $q_\mu\gtrsim 2$ fm$^{-1}$ than reported previously 
in Refs.~\cite{Piarulli:2013,Schiavilla:2019}.
The dominant isovector terms in the two-body currents convert
spin/isospin $S/T\,$=$\,0/1$ pairs into $S/T\,$=$\,1/0$ pairs, and vice versa.
Since in nuclei the correlated pair wave functions in these spin-isospin channels
are similar in shape and only differ
by a scale factor~\cite{Forest:1996}, one expects a similar scaling to occur
in the two-body isovector transition densities, defined as
\begin{equation}
\rho^{2b}(r)=\bra \psi_{pn}({}^1{S}_0) |\sum_{i<j} j_{ij,y}(q\hat{\bf x})\, \delta(r_{ij}-r)|\psi_d\ket\,,
\end{equation}
in the deuteron, and as
\begin{eqnarray}
\nonumber
  \hspace{-0.75cm}   \rho^{2b}(r)&=&\bra \psi_{3N} |\sum_{i<j} P_{ij}^{01}j_{ij,y}(q\hat{\bf x})\,\delta(r_{ij}-r)P_{ij}^{10}|\psi_{3N}\ket\\
    &+&\!\bra \psi_{3N} |\sum_{i<j} P_{ij}^{10}j_{ij,y}(q\hat{\bf x})\,\delta(r_{ij}-r)P_{ij}^{01}|\psi_{3N}\ket\ ,
\end{eqnarray}
in the trinucleons,
where $\psi_d$ and $\psi_{3N}$ are, respectively,
the deuteron and trinucleon wave functions, $\psi_{np}({}^1{S}_0)$ is the $np$ scattering
wave function in the ${}^1S_0$ channel, $j_{ij,y}$ is the $y$ component of the two-body
isovector current (including all terms up to N3LO), and $P_{ij}^{ST}$ is the projector operator over $ST$ states for the pair $ij$.
In Fig.~\ref{fig:density} we show the densities corresponding
to the NVIa$^*$ and EMN500 interactions.  They have been computed at  $q_\mu \approx 0.67$  and $3.45$ fm$^{-1}$,
and have been rescaled so as to peak at 1.  As expected,
the ratio of two-body current matrix elements
in the $d$-threshold cross section and $^3$H/$^3$He magnetic form factors
is very nearly the same. Therefore, knowledge of these matrix elements in $d$-threshold
is sufficient to predict the corresponding matrix elements in the $^3$H/$^3$He
form factors (or vice versa).  In the region of $q_\mu\gtrsim 3.5$ fm$^{-1}$,
the contribution of two-body currents in these observables is dominant. 
It is then not surprising that, if we reproduce the $d$-threshold cross section in this $q_\mu$-region,
we will also be able to reproduce the trinucleon magnetic form factors.
Incidentally, this scaling behavior also occurs in Gamow-Teller matrix elements
of two-body weak currents in light nuclei~\cite{King:2020}.

We conclude this section by making a few remarks regarding the chiral convergence and the
systematic errors generated by the truncation of the chiral expansion.
As case study, we select the EMN500 model, which provides the
best description of the trinucleon $F_M(q_\mu)$ and for which we have interactions
at increasing orders, from LO to N4LO. 
While a Bayesian analysis based on Ref.~\cite{Melendez:2019} is ongoing, for the time
being we carry out a study based on the approach by Epelbaum {\it et al.}~\cite{Epelbaum:2015}.  For the observable $F_M(q_\mu)$, we consider
two distinct expansions, one for the nuclear interaction
and one for the electromagnetic current. Therefore, we label the magnetic form factor
as $F^{i,j}_M(q_\mu)$,
where with  $i$ (N$i$LO, $i=0$ for LO) we indicate the order of the interaction
and with $j$ the order
of the electromagnetic current at which $F_M(q_\mu)$ has been computed.
To estimate the error due to the truncation of the interaction, we fix the
current at N3LO, and then use the prescription of Ref.~\cite{Epelbaum:2015}, namely
\begin{equation}
  \begin{aligned}
    \Delta F_M^{I}(q_\mu)=&\max\Big[\alpha^6\times|F_M^{0,3}(q_\mu)|,\\
    &\alpha^4\times|F_M^{1,3}(q_\mu)-F_M^{0,3}(q_\mu)|,\\
    &\alpha^3\times|F_M^{2,3}(q_\mu)-F_M^{1,3}(q_\mu)|,\\
    &\alpha^2\times|F_M^{3,3}(q_\mu)-F_M^{2,3}(q_\mu)|,\\
    &\alpha\times|F_M^{4,3}(q_\mu)-F_M^{3,3}(q_\mu)|\Big]\,.
    \end{aligned}
\end{equation}
Similarly, for estimating the uncertainty due to the truncation of the current,
we fix the order of the interaction at N4LO, and evaluate
\begin{equation}
  \begin{aligned}
  \Delta F_M^{C}(q_\mu)=&\max\Big[ \alpha^4\times|F_M^{4,0}(q_\mu)|,\\
  &\alpha^3\times|F_M^{4,1}(q_\mu)-F_M^{4,0}(q_\mu)|,\\
  &\alpha^2\times|F_M^{4,2}(q_\mu)-F_M^{4,1}(q_\mu)|,\\
  &\alpha\times|F_M^{4,3}(q_\mu)-F_M^{4,2}(q_\mu)|\Big]\,.
  \end{aligned}
\end{equation}
For semplicity, we assume that the two uncertainties are independent, and conservatively
estimate the total uncertainty as
\begin{equation}
  \Delta F_M(q_\mu)=\Delta F_M^{I}(q_\mu)+\Delta F_M^{C}(q_\mu)\,.
\end{equation}
The expansion parameter $\alpha$ is taken as
\begin{equation}
\alpha=\max\left [\frac{|{\bf q}|}{\Lambda_\chi}\,,\, \frac{m_\pi}{\Lambda_\chi}\right ]\,,
\end{equation}
where $|{\bf q}|$ is the magnitude of the three-momentum transfer
and  $\Lambda_\chi\simeq 1$ GeV.  Of course, there is some degree
of arbitrariness in the choice of $\alpha$, since, for example,
the external electromagnetic field, when it couples to a single nucleon, imparts a
momentum ${\bf q}$ to this nucleon, while, when it couples
to a pair of nucleons, will impart, on average, ${\bf q}/2$ to each
nucleon in the pair.

\begin{figure}[bth]
  \hspace*{-0.5cm}
  \includegraphics[scale=0.7]{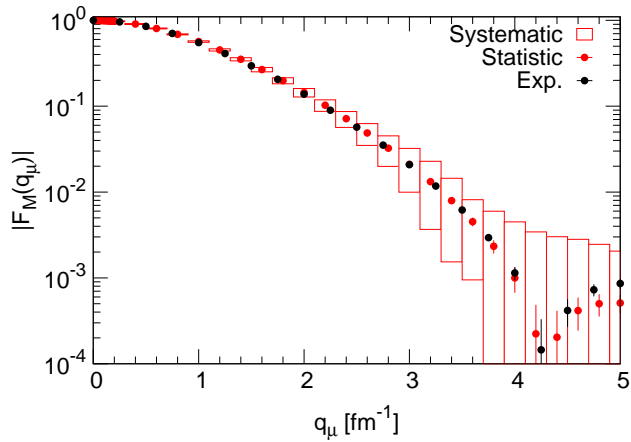}
  \caption{\label{fig:err1} The $\Het$ magnetic form factor obtained
  with the EMN500 Hamiltonian model.  The red boxes represent the uncertainties due to the truncation of the chiral expansion in the electromagnetic current and two-nucleon
 interaction (systematic). With the red bar we indicate instead the uncertainties
produced by the fit of the LECs that appear in the electromagnetic current (statistical).}
\end{figure}

In Fig.~\ref{fig:err1} we plot the results of the present analysis
for the magnetic form factor of $\Het$.  The red boxes represent
the systematic uncertainty $\Delta F_M(q_\mu)$.  The red points
are the central values and the bar represents the statistical uncertainty
due to the fitting procedure. It is immediately clear that this statistical
uncertainty is almost negligible, when compared to the systematic one.
The systematic uncertainty increases as $q_\mu$ increases, and seems
to much reduce the predictive power of the theory for $q_\mu \gtrsim3.8$ fm$^{-1}$,
albeit it should be noted that in that region the form factor has a zero. 

\begin{figure}[h]
    \hspace*{-0.5cm}
    \includegraphics[scale=0.35]{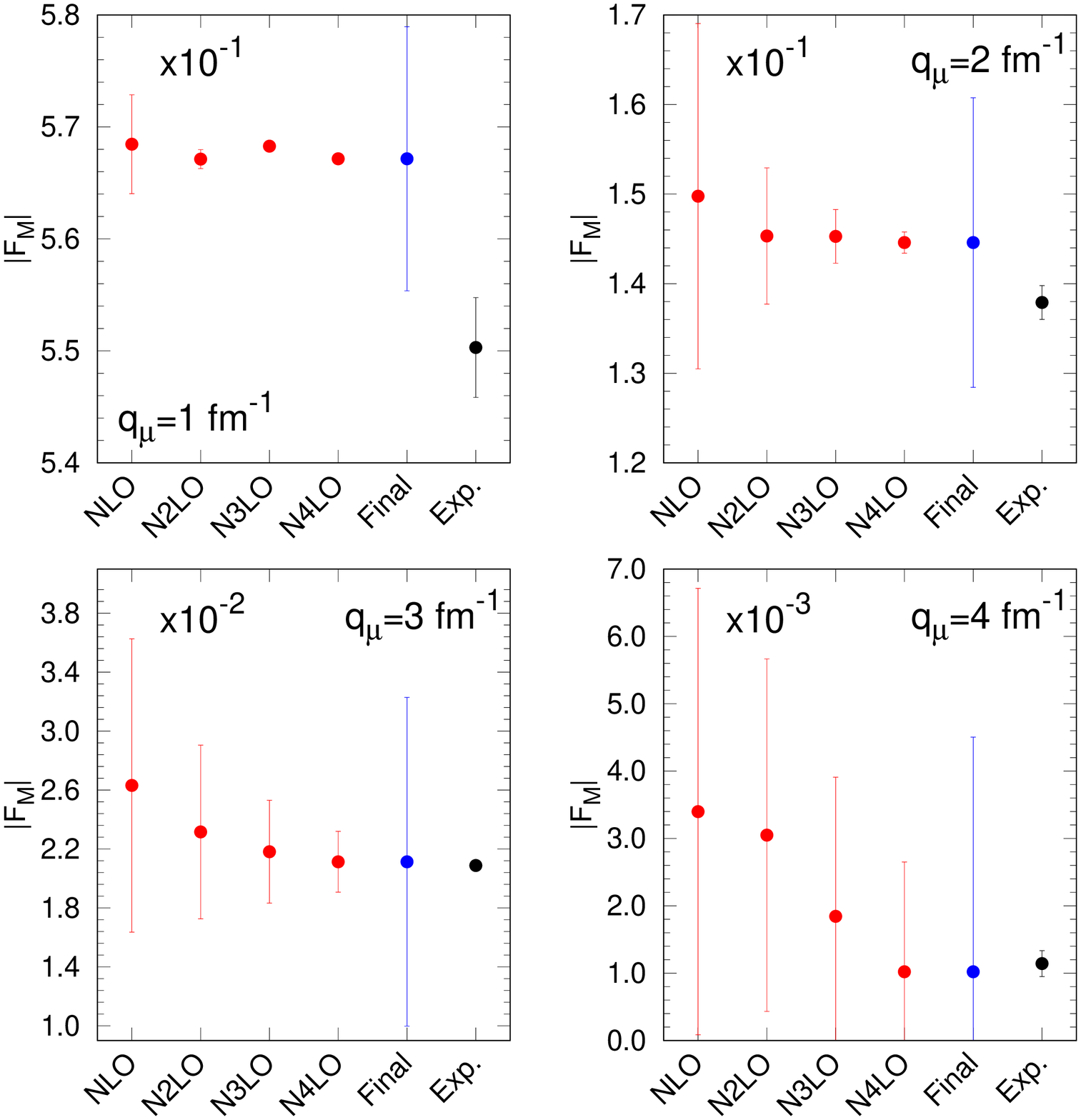}
 \caption{\label{fig:err2} Convergence results at selected values of $q_\mu$ for
the $\Het$ magnetic form factor.  They are shown order by order in the expansion of the two-nucleon interaction, but with the electromagnetic current
fixed at N3LO.  The final points (blue) represent the full calculation, which includes
in addition the systematic uncertainty due to the truncation of
the electromagnetic current.  The experimental values are also reported for comparison.}
\end{figure}

In Fig.~\ref{fig:err2} we show the result for the $\Het$ magnetic form factor
for selected values of $q_\mu$ order by order in the two-nucleon interaction,
but with the current fixed at N3LO.  The error bars on the red points represent only the systematic uncertainties $\Delta F_M^C(q_\mu)$.  For all the $q_\mu<4$ fm$^{-1}$ considered the convergence of the chiral expansion appears to be satisfactory.
At larger $q_\mu$, the chiral expansion loses its predictive power
and the systematic uncertainties dominate, as already observed in Fig.~\ref{fig:err1}.
It is also worthwhile noting that for $q_\mu=1$ and $2$ fm$^{-1}$, the point of
convergence for the expansion appears to be off the experimental value.
However, by adding the uncertainty due to the truncation of the chiral expansion in the electromagnetic current (blue points indicated by Final in Fig.~\ref{fig:err2}), the theoretical predictions become compatible with the experimental values (within 2$\sigma$ for $q_\mu=1$ fm$^{-1}$).  Similar conclusions hold for the magnetic form factors of $d$ and $\Ht$ obtained with the EMN500 interaction.

\begin{acknowledgments}
  We would like to thank Dr.~Josh Martin for pointing out the mismatch in the labeling
  of the LECs entering the NLO contact terms of the NV interactions and the LECs of the
  minimal contact current.
  A.G.~would like to thank all the LANL T2-group for the hospitality and the interesting
  discussions during his visit in Los Alamos.
  The support of the U.S.~Department of Energy, Office of Science,  Office of Nuclear Physics, under contracts
  DE-AC05-06OR23177.  Computational resources provided by the National Energy Research
  Scientific Computing Center (NERSC) are also thankfully acknowledged.
\end{acknowledgments}

\begin{thebibliography}{100}
\bibitem{NNDC}
  As reference values for the nuclear magnetic moments we used the
  NNDC catalog.
\end{thebibliography}
\end{document}